\documentclass[12pt]{iopart}
\usepackage{graphicx}

\begin{document}

\title
{Cooper pair sizes in $^{11}$Li and in superfluid nuclei: a puzzle?}

\author{K. Hagino$^1$, H. Sagawa$^2$ and P. Schuck$^{3,4}$}
\address{
$^1$
Department of Physics, Tohoku University, Sendai, 980-8578,  Japan}
\ead{hagino@nucl.phys.tohoku.ac.jp}
\address{$^2$
Center of Mathematics and Physics,  University of Aizu,
Aizu-Wakamatsu, Fukushima 965-8560,  Japan}
\address{$^3$
Institut de Physique Nucl\'eaire, CNRS, UMR8608, Orsay, F-91406,
France}
\address{$^4$
Universit\'e Paris-Sud, Orsay, F-91505, France}

\begin{abstract}
We point out a strong influence of the pairing force 
on the size of the two neutron Cooper 
pair in $^{11}$Li, and to a lesser extent also in $^6$He.  
It seems that these are quite unique situations,  
since Cooper pair sizes of stable 
superfluid nuclei are very little influenced by the intensity 
of pairing, as recently 
reported. We explore the difference between $^{11}$Li and heavier superfulid 
nuclei, and discuss reasons 
for the exceptional situation in $^{11}$Li. 
\end{abstract}

\submitto{\JPG}
%\maketitle

%\section{Introduction}

It is well known that 
pairing correlations 
enhance cross sections for two-neutron transfer reactions 
(see Refs. 
\cite{OV01,PBBVB09} for recent reviews on pair transfer). 
A few theoretical calculations have revealed that 
a spatially confined neutron pair 
({\it i.e.,} dineutron or Cooper pair) exists on the nuclear surface 
for two particles (or two holes) around a core nucleus with shell closure
\cite{BBR67,IAVF77,ZFWB80,JL83,CIMV84,ILM89}. 
The enhancement of two-neutron transfer cross sections 
has been attributed to this effect. 

Probably it is Hansen and Jonson who 
exploited the idea of dineutron correlation explicitly for 
exotic nuclei for the first time. 
They proposed the
dineutron cluster model and successfully analysed the matter radius
of $^{11}$Li \cite{HJ87}. They also predicted a large Coulomb dissociation
cross section of the $^{11}$Li nucleus. 

Recently, the dineutron correlation has attracted much attention in connection 
with 
neutron-rich nuclei, partly due to the new measurement 
for the Coulomb dissociation of $^{11}$Li \cite{N06}, 
which have shown a strong indication of the existence of a correlated 
dineutron 
in $^{11}$Li. In fact, 
many theoretical
discussions on the dineutron correlation have been taking place in recent 
years, not only in
the 2$n$ halo nuclei, $^{11}$Li and $^6$He
\cite{BBBCV01,HS05,HSCP07,BH07,HSNS09},
but also
in medium-heavy neutron-rich nuclei \cite{HTS08,IIAAK08,E07,MMS05,PSS07,AI09}
as well as in infinite neutron matter \cite{M06,MSH07,EHSS09}.
These calculations have shown that the dineutron correlation 
is enhanced in neutron-rich nuclei, although it exists also in 
stable nuclei. 

\begin{figure}
\begin{center}
\includegraphics[width=7cm,clip]{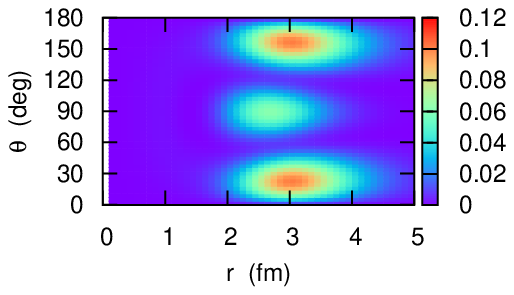}
\includegraphics[width=7cm,clip]{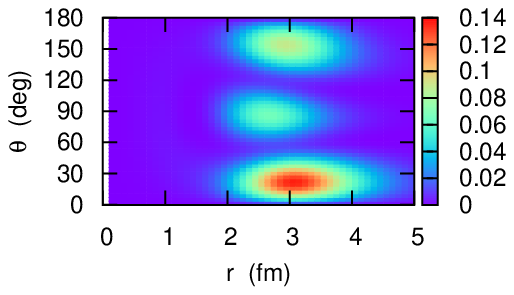}
\caption{The two particle density for $^{18}$O nucleus obtained with 
a three-body model of $^{16}$O+$n$+$n$ as a function of the distance 
of the neutron $r_1$=$r_2$=$r$ from the core nucleus and the 
opening angle between the valence neutrons, $\theta$. The left panel 
shows the two particle density obtained by including only the 
bound 1d$_{5/2}$, 2s$_{1/2}$, and 1d$_{3/2}$ single-particle 
levels in the three-body model 
calculation, while the right panel is obtained by including single particle 
levels up to $e_{\rm max}=$30 MeV and $l_{\rm max}=7$.}
\end{center}
\end{figure}

It is easy to understand why  dineutron correlations become more 
significant if the two neutrons are close to the continuum threshold.
For this purpose, we show in Fig. 1 
the result of a three-body model calculation for $^{18}$O nucleus, 
$^{16}$O+$n$+$n$, in which the valence neutrons interact with each other 
via a density-dependent contact interaction \cite{BE91}
(that is, the so called surface 
type pairing interaction)
\footnote{
We have 
also used Gogny equivalent contact forces \cite{GSGS99} and 
have reached the same conclusion on Cooper pair sizes.}.
For the $n$-$^{16}$O potential, we use the same 
Woods-Saxon potential as in Ref. \cite{IAVF77}, which has three bound 
levels,  1d$_{5/2}$, 2s$_{1/2}$, and 1d$_{3/2}$, above the $N=8$ shell 
closure. 
The left panel shows the 
two particle density, $\rho(r,r,\theta)$, as a function of 
the neutron-core distance $r$ and the opening angle between the two 
neutrons, $\theta$, obtained by including only the bound levels in the 
three-body model calculation. Here, we set $r_1=r_2=r$ for presentation purposes. 
The figure shows symmetric two peaks at $\theta\sim 0$ and 
$\theta\sim \pi$. The right panel, on the other hand, is obtained by including 
single particle levels up to $e_{\rm max}=$30 MeV and $l_{\rm max}=7$. In this case, the two peaks become 
strongly asymmetric, the peak around $\theta\sim 0$ being much more enhanced 
as compared to the other peak. This is nothing more than the manifestation of 
dineutron correlations which we shall discuss in this article. 
To obtain the spatially compact dineutron, it has been recognized that 
 the mixing of single particle levels 
of opposite parities by the pairing interaction plays 
an essential role \cite{CIMV84,PSS07}.
That is, the pairing 
interaction mixes the bound positive parity levels with a lot of 
continuum levels with negative parity. 
As 
the Fermi energy becomes smaller, 
the continuum states play a decisive role and 
the admixture of opposite parity states takes place more significantly 
in neutron-rich nuclei, leading to strong dineutron correlations. 

Although the dineutron correlation in
the $^{11}$Li nucleus has been discussed for two decades since
the publication of Hansen and Jonson,
many questions have yet to be answered, especially also for dineutron 
correlations in heavier neutron-rich nuclei. 
In this article, we discuss such open problems concerning the 
pairing properties in neutron-rich nuclei from the 
point of view of dineutron correlations. 
In particular, we shall discuss 
the local size of  dineutrons. 

%\section{The size of dineutron}

Many properties of superfluidity are related to the size of Cooper pairs 
relative to the mean interparticle distance and to the size of the
system \cite{BB05}. 
In the BCS approximation, it is well known that the size of a Cooper pair 
in infinite matter is characterized by the coherence length given by Pippard's 
relation 
\begin{equation}
\xi=\frac{\hbar^2k_F}{m^*\pi \Delta},
\label{coherence}
\end{equation}
where $k_F$ is the Fermi momentum, $m^*$ is the effective mass, and $\Delta$ is the pairing gap \cite{BB05,FW}. 
In atomic nuclei, the coherence length 
estimated with Eq. (\ref{coherence}), averaged over the volume in a 
local density approximation (LDA) 
procedure,  
is of the size of a larger nucleus.

The density dependence of the coherence length has been investigated recently 
by Matsuo for infinite nuclear and neutron matter \cite{M06}. 
It was shown that the coherence length of a $nn$ pair first 
shrinks as density decreases from the normal density, and then 
expands again after taking a minimum at around $\rho/\rho_0\sim 0.1$. 
See also Refs. \cite{LS01} for a similar behaviour for a 
$np$ pair, which has a bound state at $\rho=0$. 

In Ref. \cite{HSCP07}, we have shown that the root mean square distance 
between the valence neutrons in $^{11}$Li exhibits a qualitatively similar 
density 
dependence as the pair moves from the center of the nucleus to the nuclear 
surface and to free space 
\footnote{The size of deuteron in $^6$Li also behaves similarly \cite{IB70}. 
However, it is not clear whether the mechanism is the same (see below).}.
Moreover, subsequent Hartree-Fock-Bogoliubov calculations with 
the Gogny interaction have confirmed that the size shrinking behaviour 
exists generically also in medium-heavy 
nuclei with strong superfluidity, both in 
stable and neutron-rich nuclei\cite{PSS07}. See also Ref. \cite{PBBV08}. 

\begin{figure}
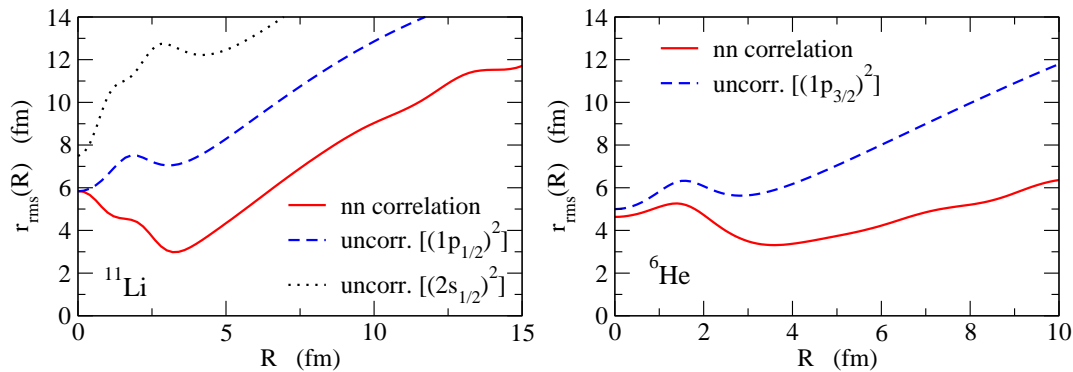

\begin{center}
\includegraphics[width=7cm,clip]{fig2a.eps}
\includegraphics[width=7cm,clip]{fig2b.eps}
\caption{(the left panel)
The root mean square distance $r_{\rm rms}$ 
for the neutron Cooper pair in $^{11}$Li as a function of
the nuclear radius $R$. The solid line shows the result of 
three-body model calculation with density dependent
contact pairing force, while the dashed and the dotted 
lines are obtained by switching
off the neutron-neutron interaction and 
assuming 
[(1p$_{1/2}$)$^2$] and [(2s$_{1/2})^2$] 
configurations,
respectively. For the uncorrelated cases, the single-particle 
potential is adjusted so that the corresponding 
single-particle energy is $-$0.15 MeV.
(the right panel)The same as the left panel, but for $^6$He nucleus. 
The dashed line corresponds to the uncorrelated pair in the 
1p$_{3/2}$ orbital at $e=-0.49$ MeV. 
}
\end{center}
\end{figure}

A question has arisen concerning what causes a small Cooper pair 
on the nuclear surface. 
As we have mentioned, in finite nuclei, the coherence length 
estimated in infinite nuclear matter, Eq. (\ref{coherence}), is about 
the nuclear size.

Very recently, the size effect has been studied 
for the $^{120}$Sn nucleus \cite{PSSB09}. 
It has turned out that the relative importance 
between pairing and size effects is totally opposite 
between $^{11}$Li and $^{120}$Sn. 
That is, the coherence length for a Cooper pair in $^{120}$Sn is affected 
very little by the pairing interaction and takes a minimum of about 2 fm on 
the surface even 
in a situation of negligible pairing gap. 
This has been seen  for other superfluid nuclei in various mass 
regions as well \cite{PSSB09}. 
 
Indeed, for the $^{11}$Li and $^6$He nuclei, 
the pairing effect seems to dominate over the 
size effect. 
This is demonstrated in Fig. 2, where we 
show 
the local coherence length of the Cooper pair in $^{11}$Li and $^6$He 
as a function of the nuclear
radius $R$ obtained with and without the $nn$ interaction. 
For the
uncorrelated calculations for $^{11}$Li, 
we consider both the [(1p$_{1/2})^2$] and [(2s$_{1/2})^2$] 
configurations 
and adjust the single-particle potential so that the 
corresponding single-particle energy is $-$0.15 MeV. 
For $^6$He, 
we consider 
an uncorrelated pair in the 1p$_{3/2}$ orbital 
at $-0.49$ MeV. 
One can see that, in the non-interacting case, the Cooper pair continuously
expands, as it gets farther away from the center of the nucleus. 
In marked contrast,
in the interacting case it becomes smaller 
going from inside to the surface before
expanding again into the free space configuration.

\begin{figure}
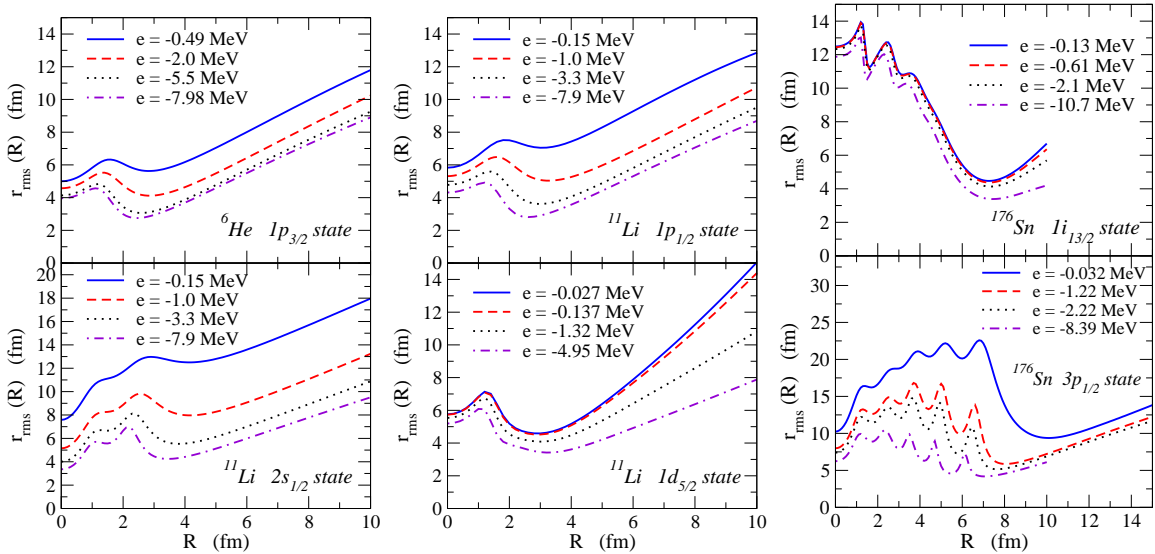

\begin{center}
\includegraphics[width=5cm,clip]{fig3a.eps}
\includegraphics[width=5cm,clip]{fig3b.eps}
\includegraphics[width=5cm,clip]{fig3c.eps}
\caption{
The root mean square distance $r_{\rm rms}$ 
for the {\it uncorrelated} 
neutron Cooper pair in $^6$He, $^{11}$Li, and 
$^{176}$Sn nuclei for various single-particle angular momenta 
and energies as indicated in the figure. }
\end{center}
\end{figure}

A reason why 
$^6$He and $^{11}$Li behave 
differently from $^{120}$Sn 
with respect to the coherence length 
may be that the neutron pairs in $^6$He and $^{11}$Li are bound much more 
weakly 
than in $^{120}$Sn. We will argue that the main reason is that the dominant 
components in 
the ground state wave function in $^6$He and $^{11}$Li are 
low angular 
momentum states with zero or one node. 
This may be inferred from the fact that the rms distance 
for uncorrelated (2s$_{1/2})^2$ and 
(1p$_{1/2})^2$ pairs in $^{11}$Li, as well as  
an uncorrelated (1p$_{3/2})^2$ pair in $^6$He, 
take a 
pronounced 
minimum when 
the binding is deep, as shown 
in Fig. 3. In this case, the behaviour of the rms distance indeed 
resembles the one for the correlated pair shown in Fig. 2. On the other hand, 
for an uncorrelated (1d$_{5/2})^2$ pair in $^{11}$Li, the rms distance 
shows a clear minimum even when the binding is extremely weak (see the middle 
lower panel in Fig. 3). 
This can be explained by the fact 
that 
a halo wave function is only connected with s- and p-waves
in the zero energy limit because of the small centrifugal 
barrier, as has been studied in Refs. \cite{HS92,RJM92}. 
For an uncorrelated pair in $^{176}$Sn, on the other hand, 
the rms distance takes a minimum both for 
(1i$_{13/2})^2$ and (3p$_{1/2})^2$ configurations even for a small 
binding energy, 
although 
the rms distance 
of the latter does not get lower than 10 fm for $e=-0.032$ MeV and 
the dependence on the single-particle energy is much
stronger in (3p$_{1/2})^2$ than in (1i$_{13/2})^2$ (see the right panels 
in Fig. 3). 
For the (3p$_{1/2})^2$ configuration, the higher nodal structure 
may cause the difference between $^{11}$Li and $^{176}$Sn. 
At any rate, $^{11}$Li and $^{6}$He seem to be very unique cases with 
respect to the influence of the pairing interaction on the value of the 
local rms distance of the Cooper pair. It would be interesting to find 
further exceptional examples of this kind in the nuclear chart. 
In general, one must conclude that besides rare cases such as $^{11}$Li 
and to a 
lesser extent $^6$He, the small rms radius of Cooper pairs in the surface of 
nuclei is essentially provoked by the size dependence of the single particle 
wave 
functions and not by pairing. The influence of the latter drops out from 
a compensation in numerator and denominator of the normalised two body wave 
function \cite{PSSB09, VSBPS09}.

\begin{figure}
\begin{center}
\includegraphics{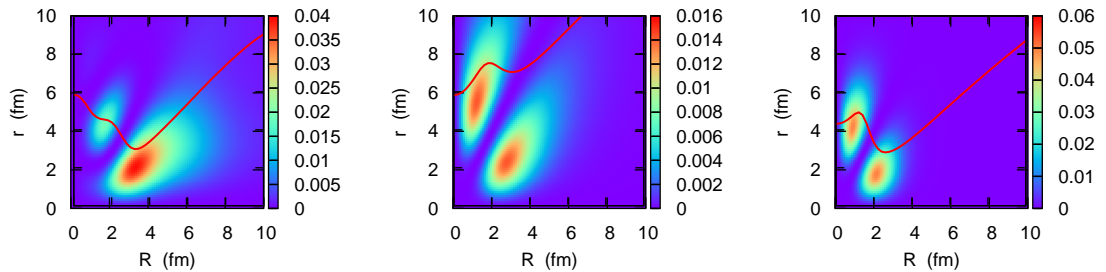}
\caption{
The square of the radial part of two-particle wave function for $^{11}$Li. 
The multiplicative factor of 
$r^2R^2$ is taken into account. 
The left, middle, and right panels correspond to the 
the correlated pair, 
the uncorrelated 
$(1p_{1/2})^2$ 
pair with the single-particle energy of $e=-0.15$ MeV, 
and the uncorrelated $(1p_{1/2})^2$ pair 
with $e=-7.9$ MeV, respectively. 
The local coherence length as a function of $R$ shown in Figs. 2 and 3 
is also plotted by the solid line. 
}
\end{center}
\end{figure}

\begin{figure}
\begin{center}
\includegraphics{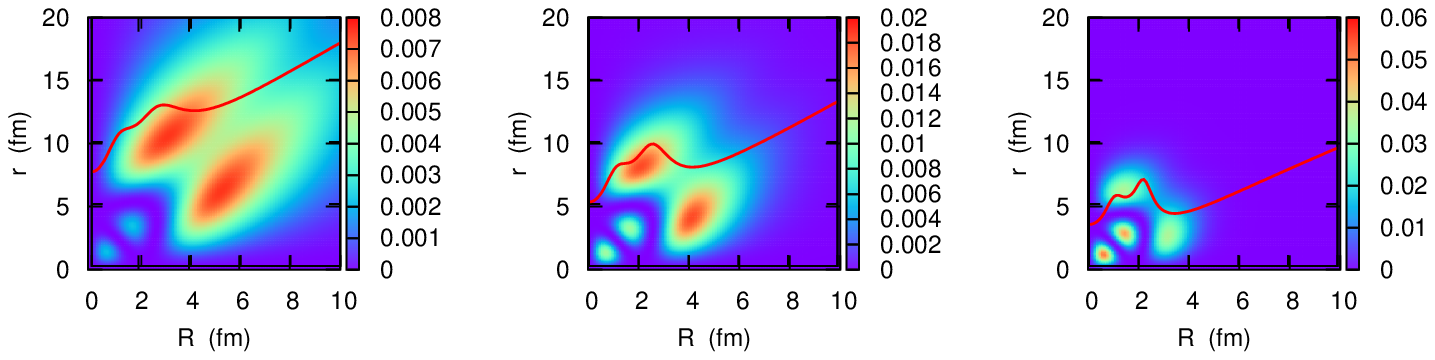}
\caption{
Same as Fig. 4, but for an uncorrelated $(2s_{1/2})^2$ pair in $^{11}$Li 
with $e=-0.15$ MeV (the left panel), $e=-1.0$ MeV (the middle panel), 
and $e=-7.9$ MeV (the right panel). }
\end{center}
\end{figure}

\begin{figure}
\begin{center}
\includegraphics{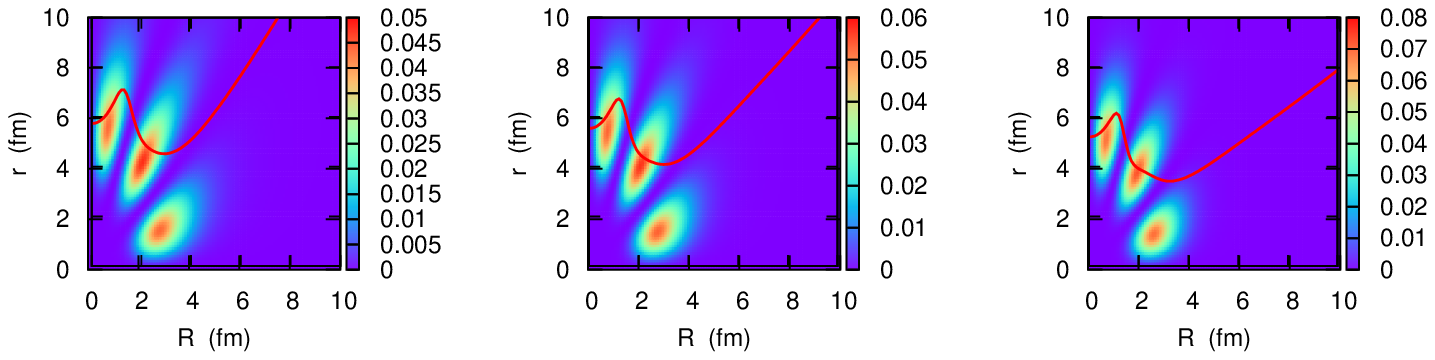}
\caption{
Same as Fig. 4, but for an uncorrelated $(1d_{5/2})^2$ pair in $^{11}$Li 
with $e=-0.137$ MeV (the left panel), $e=-1.32$ MeV (the middle panel), 
and $e=-4.95$ MeV (the right panel). }
\end{center}
\end{figure}

\begin{figure}
\begin{center}
\includegraphics{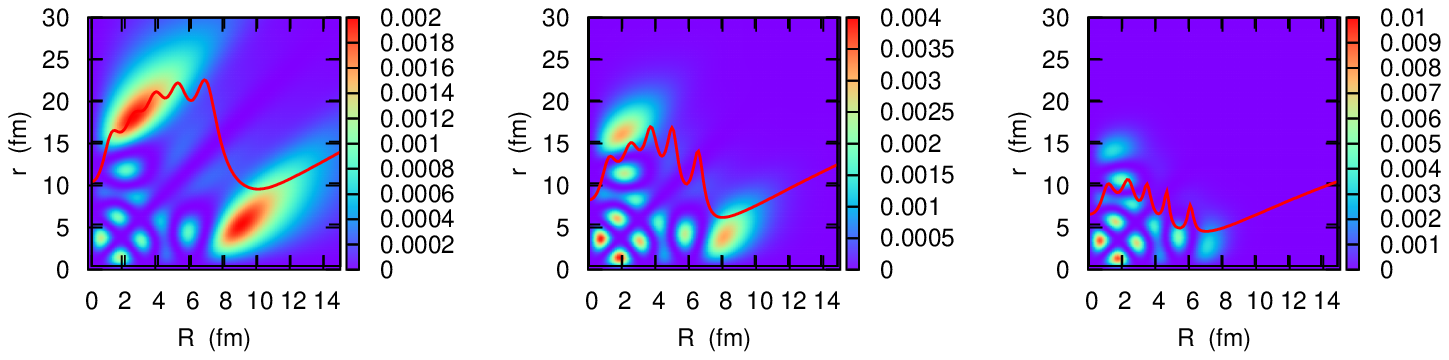}
\caption{
Same as Fig. 4, but for an uncorrelated $(3p_{1/2})^2$ pair in $^{176}$Sn 
with $e=-0.032$ MeV (the left panel), $e=-1.22$ MeV (the middle panel), 
and $e=-8.39$ MeV (the right panel). }
\end{center}
\end{figure}

Even though in general  pairing does not seem to play an important 
role in the coherence length of Cooper pairs in standard superfluid nuclei, 
one should not forget its important influence on other quantities, as 
{\it e.g.} 
the strong reduction of the moment of inertia from its classical value. 
We have already seen in Fig. 1 the strong influence of pairing interaction 
also on the density distribution of $^{18}$O. 
In Fig. 4, we demonstrate it again 
for $^{11}$Li in a different way in connection to the size of Cooper pair.  
That is, Fig. 4 shows the two dimensional plot for the square of the radial 
part of two-particle wave function, 
$\Psi(R,r)^2$, for $^{11}$Li multiplied by 
$r^2R^2$. 
The solid line denotes the local coherence length shown in Figs. 2 and 3. 
The left, the middle, and the right panels correspond to 
the correlated pair, 
the uncorrelated $(1p_{1/2})^2$ 
pair with the single-particle energy of $e=-0.15$ MeV, 
and the uncorrelated pair 
with $e=-7.9$ MeV, respectively. 
For the uncorrelated pair, there are two peaks with almost the same 
height. One of the peaks is located at small $r$, and this peak is 
monitored when 
$R$ is increased from $R=0$, leading to the minimum in the local coherence 
length 
for the uncorrelated pair with $e=-7.9$ MeV. For the uncorrelated pair 
with $e=-0.15$ MeV, both 
peaks contribute to the local coherence length at around $R\sim 3$ fm, 
and the behaviour of 
rms distance appears more complex. 
For the correlated pair, on the other hand, 
the peak with larger $r$ is much smaller than the peak 
with smaller $r$, due to the strong pairing effect. 
Therefore, it seems to be a general effect, not depending on a particular 
nucleus.
For comparisons, we also show the square of the two-particle wave functions 
for the $s$ and $d$ waves in $^{11}$Li and $p$ wave in $^{176}$Sn 
in Figs. 5,6  and 7, respectively. 

The size of Cooper pairs in  resonantly interacting atomic gases has been 
measured using 
radio-frequency spectroscopy \cite{SSSK08}. 
In nuclear physics, 
a two-neutron transfer reaction has been considered to be a good probe of 
pairing correlation, although 
it would be extremely difficult to measure the size of Cooper pairs 
directly. 

%\section{Conclusions}

In summary, we argued that the very small size of Cooper 
pairs of about 2 fm, which have recently been pointed out in several 
works to 
exist on the surface of finite nuclei, may be of radically 
different origin in various nuclei. 
Actually it seems that in most cases 
this small size of Cooper pairs on the nuclear surface 
has nothing to do with 
enhanced pairing correlations on the surface of nuclei but rather is 
a consequence of the finiteness of the single-particle wave functions \cite{PSSB09}. 
On the contrary, and this seems to be a quite unique and exceptional 
situation, in $^{11}$Li and to a lesser extent also in $^6$He, the 
Cooper pair size seems to be strongly influenced by the pairing interaction. 
This stems from the fact that the single-particle wave 
functions mostly involved are 2$s$ and 1$p$ states with very small binding. 
In that  case ($l\le1$), the centrifugal barrier is
  very low and the single-particle 
wave functions can spread out very far (to infinity in the zero energy limit
\cite{RJM92}), and make a halo structure \cite{HS92}. 
It would be important to exploit 
this unique situation of $^{11}$Li and study the influence and structure of 
the effective $nn$ force in much detail. Analysis of an ongoing experimental 
work in $^{11}$Li \cite{N09} is therefore most important. 
Investigations of whether further similar cases to $^{11}$Li and $^6$He exist 
in the nuclear chart for heavier nuclei may  be very relevant. 

\ack

We thank N. Pillet, N. Sandulescu, J. Margueron, and M. Matsuo for useful 
discussions. 
This work was supported by the Japanese
Ministry of Education, Culture, Sports, Science and Technology
by a Grant-in-Aid for Scientific Research under
the program numbers (C) 20540277 and 19740115.


\section*{References}
\begin{thebibliography}{99}

\bibitem{OV01}von Oertzen W and Vitturi A 2001 {\it Rep. Prog. Phys.} 64 1247.

\bibitem{PBBVB09}
Potel G, Bayman B F, Barranco F, Vigezzi E, and Broglia R A, arXiv:0906.4298 [nucl-th]. 

\bibitem{BBR67}
Bertsch G F, Broglia R A, Riedel C 1967 {\it Nucl. Phys.} A91 123. 

\bibitem{IAVF77}Ibarra R H, Austern N, Vallieres M, and Feng D H 1977 {\it Nucl. Phys.} 
A288 397. 

\bibitem{ZFWB80}Zhukov M V, Feng Da Hsuan, Wu C, and Bang J 1980 
{\it Physica Scripta} 22 426. 

\bibitem{JL83}Janouch F A, Liotta R J 1983 {\it Phys. Rev.} C27 896. 

\bibitem{CIMV84}Catara F, Insolia A, Maglione E, and Vitturi A 1984 {\it Phys. Rev.} C29 1091. 

\bibitem{ILM89}Insolia A, Liotta R J and Maglione E 1989 {\it J. of Phys. G} 
15 1249. 

\bibitem{HJ87}
Hansen P G and Jonson B 1987 {\it Europhys. Lett.} 4 409. 

\bibitem{N06}Nakamura T {\it et al.} 2006 
%A.M. Vinodkumar, T. Sugimoto, N. Aoi, H. Baba, D. Bazin, N. Fukuda,
%T. Gomi, H. Hasegawa, N. Imai, M. Ishihara, T. Kobayashi, Y. Kondo,
%T. Kubo, M. Miura, T. Motobayashi, H. Otsu, A. Saito, H. Sakurai,
%S. Shimoura, K. Watanabe, Y.X. Watanabe, T. Yakushiji,
%Y. Yanagisawa, and K. Yoneda,
{\it Phys. Rev. Lett.} 96 252502.

\bibitem{BBBCV01}Barranco F, Bortignon P F, Broglia R A,
Colo G, and Vigezzi E 2001 {\it Eur. Phys. J.} A11 385.

\bibitem{HS05}
Hagino K and Sagawa H 2005 {\it 
Phys. Rev.} C72 044321;
2007 {\it Phys. Rev.} C75 021301.

\bibitem{HSCP07}Hagino K, Sagawa H, Carbonell J, and Schuck P,
2007 {\it Phys. Rev. Lett.}  99 022506.

\bibitem{BH07}Bertulani C A and Hussein M S 2007 
{\it Phys. Rev.} C76 051602.

\bibitem{HSNS09}Hagino K, Sagawa H, Nakamura T and Shimoura S 2009 
{\it Phys. Rev.} C80 031301(R). 

\bibitem{HTS08}Hagino K, Takahashi N and Sagawa H 2008 
{\it Phys. Rev.} C77 054317.

\bibitem{IIAAK08}Itagaki N, Ito M, Arai K, Aoyama S, and Kokalova T 2008 
{\it Phys. Rev.} C78 017306. 

\bibitem{E07}Kanada-En'yo Y 2007 {\it Phys. Rev.} C76 044323. 

\bibitem{MMS05} Matsuo M, Mizuyama K and Serizawa Y 2005 
{\it Phys. Rev.} C71 064326.

\bibitem{PSS07}Pillet N, Sandulescu N, and Schuck P 2007 
{\it Phys. Rev.} C76 024310.

\bibitem{AI09}Aoyama S and Itagaki N 2009 {\it Phys. Rev.} C80 021304. 

\bibitem{M06}Matsuo M 2006 {\it Phys. Rev.} C73 044309.

\bibitem{MSH07}Margueron J, Sagawa H, and Hagino K 2007 
{\it Phys. Rev.} C76 064316.

\bibitem{EHSS09}Kanada-En'yo Y, Hinohara N, Suhara T, and Schuck P 2009 
{\it Phys. Rev.} C79 054305. 

\bibitem{BE91}
Bertsch G F and  Esbensen H 1991 {\it Ann. Phys. (N.Y.)} 209 327. 

\bibitem{GSGS99}Garrido E, Sarriguren P, Moya de Guerra E, and 
Schuck P 1999 {\it Phys. Rev.} C60 064312. 

\bibitem{BB05}Brink D M and Broglia R A 2005 
{\it Nuclear Superfluidity} (Cambridge: Cambridge University Press). 

\bibitem{FW}Fetter A L and Waleck J D 1971 
{\it Quantum Theory of Many-Particle Systems} 
(McGraw-Hill). 

\bibitem{LS01}Lombardo U and Schuck P 2001 {\it Phys. Rev. C} 63 038201. 

\bibitem{IB70}Itonaga K and Bando H 1970 {\it Prog. Theo. Phys.} 44 1232. 

\bibitem{PBBV08}Pastore A, Barranco F, Broglia R A, and Vigezzi E 2008 
{\it Phys. Rev. C} 78 024315. 

\bibitem{PSSB09}Pillet N, Sandulescu N, Schuck P, and Berger J -F 2009 
{\it preprint}. 

\bibitem{HS92} Sagawa H,  1992 {\it  Phys. Lett. B} 286, 7.

\bibitem{RJM92}Riisager K, Jensen A S, and Moller P 1992 {\it Nucl. Phys. A} 548 393. 

\bibitem{VSBPS09}Vinas X, Schuck P, Berger J F, Pillet N, and Sandulescu N, to be published.

\bibitem{SSSK08}Schunck C H, Shin Y, Schirotzek  A, and Ketterle W 
2008 {\it Nature} 454 739. 

\bibitem{N09}Nakamura T, private communication. 

\end{thebibliography}
\end{document}